\documentclass[preprint,12pt]{JHEP3}
\JHEPspecialurl{http://jhep.sissa.it/JOURNAL/JHEP3.tar.gz}
\usepackage{epsfig,multicol}

\title{Systematics of Moduli Stabilization, Inflationary Dynamics and Power Spectrum}

\author{R. Holman and Jimmy A. Hutasoit\\
Dept. of Physics, Carnegie Mellon University, Pittsburgh PA 15213
USA\\ E-mail: \email{rh4a@andrew.cmu.edu},
\email{jhutasoi@andrew.cmu.edu}}

\abstract{We study the scalar sector of type IIB superstring theory
compactified on Calabi-Yau orientifolds as a place to find a
mechanism of inflation in the early universe. In the large volume
limit, one can stabilize the moduli in stages using perturbative
method. We relate the systematics of moduli stabilization with
methods to reduce the number of possible inflatons, which in turn
lead to a simpler inflation analysis. Calculating the
order-of-magnitude of terms in the equation of motion, we show that
the methods are in fact valid. We then give the examples where these
methods are used in the literature. We also show that there are
effects of non-inflaton scalar fields on the scalar power spectrum.
For one of the two methods, these effects can be observed with the
current precision in experiments, while for the other method, the
effects might never be observable.}

\keywords{Moduli stabilization, Inflation}

\preprint{hep-th/0606089}

\begin{document}

\section{Introduction}

The use of both RR and NS-NS fluxes to generate potentials for the
moduli appearing in Calabi-Yau compactifications of Type IIB string
theory \cite{gkp,grana} has breathed new vigor into attempts to find
inflation in the effective 4-D field theory associated with such
compactifications. The generic expectation is that the potentials
for the moduli fields could be flat enough to allow for a phase of
slow-roll inflation for at least one, if not more, moduli. This
expectation has been borne out in a number of examples
\cite{racetrack,axinfle,others,kahler}.

The idea of modular inflation from string theory has been around for
some time. The earlier attempts on modular cosmology concentrated
on potentials  for the dilaton. However, a detailed study
\cite{dilatoninflation} of general properties of these
potentials shows that they are either of the runaway type or too steep for
inflation.

One of the difficulties in trying to find inflationary regimes for these potentials
is that typically, more than one field will participate in the slow-roll phase.
As an example, in \cite{racetrack}, four fields were relevant to generating inflation.
Following such a system is a complex task, and it is not unreasonable to ask whether
there are ways to reduce the number of inflaton fields that need to be tracked.
In this paper, we will list two ways of simplifying the analysis by reducing the number of
possible inflatons. These
methods have strong ties to the systematics of moduli stabilization
in the large volume scheme that were discussed in Ref.~\cite{vijay}.

We will start by reviewing
the scalar sector of type IIB superstring compactified on a large
volume Calabi-Yau orientifolds. Following Ref.~\cite{vijay}, we will show
that on the large volume limit, the potential for the moduli will have a scale
hierarchy. We then exploit this hierarchy to approach the problem
of moduli stabilization in several stages. It turns out that this
hierarchy also allows us to integrate out some of the moduli from
the theory and simplify the analysis for inflation.

The existence of more than one modulus in the problem can also
influence the power spectrum of metric perturbations observed in the
CMB. If these moduli have not yet settled into their minima during
the inflationary phase, their oscillations about these minima could
imprint itself into the power spectrum \cite{transplanckianme}.
Given the new WMAP \cite{wmap} results, we may be able to place
bounds on how quickly these fields had to have reached their minima.
Once we establish the hierarchical scale structure alluded to in the
previous paragraph, we estimate the size of these effects. We find
that in some cases, the effect could be detectable.

\section{Review of the Scalar Sector of Type IIB Superstring Theory}

Type IIB superstring theory compactified on Calabi-Yau orientifolds
$M$ yields the following four dimensional effective theory:
\begin{equation}
{\cal L} = \int d^{4}x \sqrt{-g} \left(G_{\alpha \bar{\beta}}
\partial_{\mu} \Phi^{\alpha} \partial^{\mu} \Phi^{\bar{\beta}} + V
\right),
\end{equation}
where $\alpha$, $\beta$ run over all moduli,  $G_{\alpha \bar{\beta}} = \partial_{\alpha}
\partial_{\bar{\beta}} K$ is the K\"{a}hler metric on the moduli space, and where $K$ is the K\"{a}hler potential, including the $\alpha'$ corrections \cite{becker}:
\begin{equation}
K = - \log \left[-i \int_{M} \Omega \wedge \bar{\Omega} \right] -
\log \left[-i (\tau - \bar{\tau}) \right] - 2 \log \left[\,
\frac{\xi}{2} \left( \frac{- i (\tau - \bar{\tau})}{2} \right)^{3/2}
+ e^{-3 \phi_{0} /2} \, {\cal V} \, \right] \label{eq:kahler}.
\end{equation}
Here $\tau$ is the axion-dilaton field, $\Omega$ is the (3,0)-form of the
Calabi-Yau, $\cal V $ is the classical volume of $M$ in units of
$l_{s} = (2\pi) \sqrt{\alpha'}$, and $\xi = - {\zeta(3)
\chi(M)}\slash ({2(2\pi)^{3}})$. We require that $\xi
> 0$, or $h^{2,1} > h^{1,1}$.

The scalar potential is given by:
\begin{equation}
V = e^{K} \left(G^{\alpha \bar{\beta}} D_{\alpha} W \bar{D_{\beta}}
\bar{W} - 3 {\left| W \right|}^{2} \right),
\end{equation}
where the superpotential $W$ is
\begin{equation}
 W = \int_{M} G_{3} \wedge \Omega + \sum_{i} A_{i} \, e^{ia_{i}\rho_{i}} \label{eq:superpotential}.
\end{equation}
The first term is the Gukov-Vafa-Witten term \cite{GVW} and the
second one is the non-perturbative part due to D3-brane instantons
\cite{inst} or gaugino condensation from wrapped D7-branes (see
\cite{gaugino} and references therein). Here $G_{3} = F_{3} - \tau
H_{3}$, with $F_{3}$ and $H_{3}$ are RR and NS-NS 3-form fluxes,
respectively, $A_{i}$ is a one-loop determinant and $a_{i} = {2
\pi}\slash {N}$, with $N$ is a positive integer. Also,  $\rho_{i}
\equiv b_{i} + i \tau_{i}$ is the complexified K\"{a}hler modulus
consisting of the four-cycle modulus $\tau_{i}$:
\begin{equation}
\tau_{i} = \partial_{t^{i}} {\cal V} = \frac{1}{2} D_{ijk} \, t^{j}
t^{k},
\end{equation}
and the axion $b_i$. The $t^{i}$ measures the area of two-cycle,
$D_{ijk}$ are the triple intersection numbers of the divisor basis
\cite{denef} and the classical volume is expressed as
\begin{equation}
{\cal V} = \frac{1}{6} D_{ijk} \, t^{i} t^{j} t^{k}.
\end{equation}

Equations (\ref{eq:kahler}) and (\ref{eq:superpotential}) completely
specify the theory and the problem of moduli stabilization becomes the
problem of finding solutions to $\partial_{\alpha} V = 0$.
However, visualizing the full potential and finding
its minima is a difficult task as is using the full potential to look for inflationary phases. To do this, we would have to solve the following equations of motion in an FRW
background
\begin{eqnarray}
\ddot{\Phi}^{A} + 3 H \dot{\Phi}^{A} + {\Gamma^{A}}_{BC}
\dot{\Phi}^{B} \dot{\Phi}^{C} + 2 G^{AB} V_{,B}= 0 \label{eq:eom},\\
3 H^{2} = G_{AB} \dot{\Phi}^{A} \dot{\Phi}^{B} + V,
\end{eqnarray}
where ${\Gamma^{A}}_{BC}$ is the connection on the metric of the
moduli space\footnote{Capitalized Roman letters denote {\it real} scalar fields as opposed to the complex ones indicated by the Greek indices.}. The slow-roll condition is $\epsilon
<< 1$ (for more details, see Appendix \ref{section:slow-roll}),
where the slow-roll parameter is given by
\begin{equation}
\epsilon = \frac{G_{AB} \dot{\Phi}^{A} \dot{\Phi}^{B}}{H^{2}}.
\end{equation}

As discussed in the previous section, it is almost impossible to deal with the plethora of moduli that appear in these compactifications, at least as far as inflationary dynamics is concerned. What we need is  a controlled way to ``freeze'' some of these into place at their minima, while leaving a subset of them free to induce an inflationary state for the requisite amount of time.

Thus, we want to somehow {\em consistently} decouple
some fields, collectively labeled $\{\psi^{A}\}$, from the inflationary dynamics by putting
them at the minima of the potential and let the rest of the field
$\{\phi^{A}\}$ be the inflatons, i.e.:
\begin{eqnarray}
\ddot{\phi}^{A} + 3 H \dot{\phi}^{A} + {\Gamma^{A}}_{BC}
\dot{\phi}^{B} \dot{\phi}^{C} + 2 G^{AB} V_{,B}= 0, \label{equation:eom_of_inflatons}\\
3 H^{2} = G_{AB} \dot{\phi}^{A} \dot{\phi}^{B} + V.
\end{eqnarray}
In this scenario, the slow-roll condition now becomes
\begin{equation}
\epsilon = \frac{G_{AB} \dot{\phi}^{A} \dot{\phi}^{B}}{H^{2}} << 1.
\end{equation}

The problem in doing this is that there is no reason to expect that a given choice of $\{\psi^{A}\}$
will work. In general, the solution
$\{{\psi^{A}}_{\rm min}\}$ to $\partial_{\psi^{A}} V = 0$ will be
$\{\phi^{A}\}$-dependent. If ${\psi^{A}}_{\rm min} = {\psi^{A}}_{\rm
min}\left(\{\phi^{B}\}\right)$ then
\begin{equation}
{\dot{\psi}^{A}}_{\rm min} = \frac {\partial {\psi^{A}}_{\rm
min}}{\partial {\phi^{B}}} \dot{\phi}^{B} \ne 0.
\end{equation}
Thus, a careless choice of $\{\psi^{A}\}$ could give the false
impression that the slow-roll parameter is small, i.e. inflation is
occurring, when in reality $\epsilon$ might not be small.

Furthermore, even if $\dot{\psi}^{A} = 0$, there is the possibility
that $\ddot{\psi}^{A}$ will feed on the last two terms of the
equation of motion (\ref{eq:eom}), such that on a later time
$\dot{\psi}^{A}$ will deviate significantly from $0$.

A valid choice of $\{\psi^{A}\}$ should not have the above problems.
We will show that the systematics of moduli stabilization in the large
volume scenario is strongly related to finding the valid choice of
$\{\psi^{A}\}$.

\section{Systematics of Moduli Stabilization in the Large Volume Limit}
Following \cite{vijay}, we will be working on the large volume
limit, which is defined as the limit where all $\tau_{i} \rightarrow
\infty$ except one, which we denote by $\tau_{s}$, with  $\tau_s
\sim \ln \cal V$. In this limit, the potential becomes
\begin{eqnarray}
V \equiv & & e^{K} \left(\, G^{a \bar{b}} D_{a} W \bar{D_{b}} \bar{W} + G^{\tau \bar{\tau}} D_{\tau} W \bar{D_{\tau}} \bar{W} \, \right)   \nonumber \\
&+& \left( e^{K} \frac{\xi}{2 {\cal V}} \left(W \bar{D_{\tau}}
\bar{W} + \bar{W} D_{\tau} W \right) + V_{\alpha'} + V_{\rm np1} +
V_{\rm
np2}  \right) \nonumber \\
&+& \left(V_{\rm supp1} + V_{\rm supp2} + V_{\rm supp3} + V_{\rm
supp4} \right),
\end{eqnarray}
where
\begin{eqnarray}
V_{\alpha'} &=& 3 \xi e^{K} \frac{\xi^{2} + 7 \xi {\cal V}
+ {\cal V}^2}{({\cal V} - \xi) (2 {\cal V} + \xi)^{2}} \, |W|^{2}, \nonumber \\
V_{\rm np1} &=& e^{K} G^{\rho_{s} \bar{\rho_{s}}} a_{s}^{2}
|A_{s}|^{2} e^{-2 a_{s} \tau_{s}}, \nonumber \\
V_{\rm np2} &=& e^{K} G^{\rho_{s} \bar{\rho_{l}}} i a_{s} (A_{s}
e^{i a_s \rho_{s}} \bar{W} \partial_{\bar{\rho_{l}}} K - \bar{A_{s}}
e^{- i a_s \bar{\rho_{s}}} W \partial_{\rho_{l}} \bar{K}),  \nonumber \\
V_{\rm supp1} &=& e^{K} \,G^{\rho_{l} \bar{\rho_{m}}} \left(a_{l}
A_{l} a_{m} \bar{A_{m}} e^{i (a_{l} \rho_{l} - a_{m}
\bar{\rho_{m}})} \right) \nonumber \\
V_{\rm supp2} &=& e^{K}  i \, G^{\rho_{l} \bar{\rho_{m}}}
\left(a_{l} A_{l} e^{i a_{l} \rho_{l}} \bar{W}
\partial_{\bar{\rho_{m}}} K - a_{m} \bar{A_{m}} e^{-i a_{m} \bar{\rho_{m}}}
W \partial_{\rho_{l}}K \right) \nonumber \\
V_{\rm supp3} &=& e^{K} G^{\rho_{l} \bar{\rho_{s}}} \, 2 \, {\rm
Re}\left[a_{l} A_{l} a_{s} \bar{A_{s}} e^{i (a_{l} \rho_{l} - a_{s}
\bar{\rho_{s}})} \right], \nonumber \\
V_{\rm supp4} &=& e^{K}  i \, G^{\rho_{l} \bar{\rho_{s}}} a_{l}
(A_{l} e^{i a_l \rho_{l}} \bar{W}
\partial_{\bar{\rho_{s}}} K - \bar{A_{l}} e^{- i a_l \bar{\rho_{l}}}
W \partial_{\rho_{s}} \bar{K}).
\end{eqnarray}
The indices $a$, $b$ run over the complex structure moduli, and the
K\"{a}hler moduli indices $l, m \ne s$. 

For simplicity, let us assume that all $a_i$'s are of ${\cal O}(1)$,
which corresponds to the gauge rank $N \lessapprox 10$ (we will
discuss the case for smaller $a_i$ in the last section). The first
term is positive definite and of ${\cal O}({\cal V}^{-2})$, the
second is of ${\cal O}({\cal V}^{-3})$, and the third is ${\cal
O}({\cal V}^{-2\slash 3}e^{-{\cal V}^{2/3}})$.

The hierarchy
between terms in the potential allows us to approach the problem of
moduli stabilization in three stages perturbatively using the inverse volume ${\cal V}^{-1}$ as our expansion parameter. First, we will stabilize the axion-dilaton and complex structure moduli
$\{\Phi_{I}\}$ by minimizing the leading term in the potential.
Next, by including the second term, we will stabilize $b_{s},
\tau_{s}$, and ${\cal V}$, which we denote collectively by $\{\Phi_{II}\}$.
Lastly, we will stabilize the rest of the K\"{a}hler moduli
$\{\Phi_{III}\}$ by including the exponentially-suppressed last
term.

Before we continue the discussion, let us remark that the potential above can be uplifted by adding anti-D3-branes \cite{kklt} or
by using the supersymmetric D-terms \cite{d-term}. The uplift potential can be written as
\begin{equation}
V_{\rm uplift} = \sum_i \frac{\epsilon_{\rm uplift}^{(i)}}{\tau_i^{\gamma}},
\end{equation}
where $\epsilon_{\rm uplift}^{(i)} \geq 0$ and $\gamma = 2, 3$ for the case of \cite{kklt} and \cite{d-term}, respectively. $\epsilon_{\rm uplift}^{(i)} $ will be tuned such that the uplifting term will not dominate over other terms. The reason is that dominant uplifting term will result in a runaway-type potential. Therefore, in the following discussion, we will not mention the uplifting term again as the results will remain the same.

Let $\Phi_{I} = {\Phi_{I}}_{0}$ correspond to the minimum of
\begin{equation}
e^{K} \left(G^{a \bar{b}} D_{a} W \bar{D_{b}} \bar{W} + G^{\tau
\bar{\tau}}D_{\tau} W \bar{D_{\tau}} \bar{W} \right).
\end{equation}
Since this is positive definite, this means that
$\{{\Phi_{I}}_{0}\}$ is the solution to $D_{a} W = D_{\tau} W = 0$.
We can evaluate the GVW superpotential and the first two terms of the K\"{a}hler
potential at $\{{\Phi_{I}}\} = \{{\Phi_{I}}_{0}\}$; write these values as $W_{0}$ and $K_{cs}$, respectively.

Next, let us include the ${\cal O}({\cal V}^{-3})$ terms
in the potential. Substituting ${\Phi_{I}} = {\Phi_{I}}_{0} +
{{\Phi_{I}}_{1}}\slash{\cal V}$, where $\{{\Phi_{I}}_{1}\}$ can
depend on $\{\Phi_{II}\}$, gives us
\begin{eqnarray}
V &=& \frac{e^{K_{cs}}}{{\cal V}^4} \sum {{\cal
F}_1}({\Phi_{I}}_{0}) \, {{\Phi_{I}}_{1}}^2  + V_{\alpha'} + V_{\rm
np1} + V_{\rm np2}, \nonumber \\
&\approx& V_{\alpha'} + V_{\rm np1} + V_{\rm np2} \label{eq:stage_two},
\end{eqnarray}
where now
\begin{eqnarray}
V_{\alpha'} &\sim& \frac{\xi}{{\cal V}^{3}} \, e^{K_{cs}}
|W_{0}|^{2},\nonumber \\
V_{\rm np1} &\sim& \frac{(- D_{ssj} t^{j}) a_{s}^{2} |A_{s}|^{2}
e^{-2 a_{s}
\tau_{s}} e^{K_{cs}}}{{\cal V}}, \nonumber \\
V'_{\rm np2} &=& \frac{e^{K_{cs}}}{{\cal V}^{2}} G^{\rho_{s}
\bar{\rho_{l}}} i a_{s} (A_{s} e^{i a_s \rho_{s}} \bar{W_0}
\partial_{\bar{\rho_{l}}} K - \bar{A_{s}} e^{- i a_s
\bar{\rho_{s}}} W_0 \partial_{\rho_{l}} \bar{K}).
\end{eqnarray}
The minimum of the potential up to ${\cal O}({\cal V}^{-3})$
is then given by
\begin{eqnarray}
{\Phi_{I}}_{\rm min} &=& {\Phi_{I}}_{0},
\nonumber \\
{\Phi_{II}}_{\rm min} &=& {\Phi_{II}}_{0},
\label{eq:solution_stage_two}
\end{eqnarray}
with $\Phi_{II} = {\Phi_{II}}_{0}$ the solution to
$\partial_{\Phi_{II}} \left(V_{\alpha'} + V_{\rm np1} + V_{\rm np2} \right) = 0$. Of course, we can continue this
systematically order by order. Let the minimum value of the
potential at the end of this second stage be $V_0$.

Now, let us include the exponentially suppressed part of the
potential. Substituting\footnote{Since the volume is also a modulus
we stabilized in the second stage, instead of using the full
solution $\cal V$, we use the leading term ${\cal V}_0$ in our
perturbation.}
\begin{eqnarray}
\Phi_{I} &=& {\Phi_{I}}_{0} + \cdots + \frac{{\Phi_{I}}_{2}}{ {{\cal
V}_0}^{2/3} e^{{\cal V}^{2/3}_0}},
\nonumber \\
\Phi_{II} &=& {\Phi_{II}}_{0} + \cdots + {\Phi_{II}}_{2}
\frac{{{\cal V}_0}^{1/3}}{e^{{\cal V}^{2/3}_0}},
\end{eqnarray}
where ${\Phi_{I}}_{2}$ and ${\Phi_{II}}_{2}$ can be dependent on
$\{\Phi_{III}\}$ we get
\begin{eqnarray}
V &=& V_0 + \frac{e^{K_{cs}}}{{{\cal V}_0}^{4/3} e^{2{\cal V}^{2/3}_0}}
\sum {\cal F}_2({\Phi_{I}}_{0},{\Phi_{II}}_{0}) {{\Phi_{I}}_{2}}^2 +
\frac{e^{K_{cs}}}{{{\cal V}_0}^{4/3} e^{2{\cal V}^{2/3}_0}} \sum {\cal
F}_3({\Phi_{I}}_{0},{\Phi_{II}}_{0}) {{\Phi_{II}}_{2}}^2 + V_{\rm
supp1} + V_{\rm supp2}, \nonumber \\
&\approx& V_0 + V_{\rm supp1} + V_{\rm supp2},
\label{eq:stage_three}
\end{eqnarray}
where $V_{\rm supp1}$ and $V_{\rm supp2}$ are independent of
${\Phi_{I}}_{2}$ and ${\Phi_{II}}_{2}$. Thus, we have the minimum of
the full potential at
\begin{eqnarray}
{\Phi_{I}}_{\rm min} &=& {\Phi_{I}}_{0} + \cdots + {\cal
O}(\frac{1}{{{\cal V}_0}^{2/3} e^{{\cal V}^{2/3}_0}}),
\nonumber \\
{\Phi_{II}}_{\rm min} &=& {\Phi_{II}}_{0} + \cdots + {\cal
O}(\frac{{{\cal V}_0}^{1/3}}{e^{{\cal
V}^{2/3}_0}}), \nonumber \\
{\Phi_{III}}_{\rm min} &=& {\Phi_{III}}_{0},
\label{eq:solution_stage_three}
\end{eqnarray}
where ${\Phi_{III}}_{0}$ is the solution to $\partial_{\Phi_{III}}
\left(V_{\rm supp1} + V_{\rm supp2}\right) = 0$.

Neglecting volume suppressed terms, solution
(\ref{eq:solution_stage_three}) can be written as
\begin{equation}
{\Phi_{I}}_{\rm min} = {\Phi_{I}}_{0}, \, \, \, {\Phi_{II}}_{\rm
min} = {\Phi_{II}}_{0}, \, \, \, {\Phi_{III}}_{\rm min}= {\Phi_{III}}_{0}.
\end{equation}
This means that solving $\partial_{\alpha} V = 0$ perturbatively can
also be understood as a (Wilsonian) effective field theory approach: stabilize
$\{\Phi_{I}\}$ by using only the leading term of the potential,
integrate $\{\Phi_{I}\}$ out, stabilize $\{\Phi_{II}\}$ with the
${\cal O}({\cal V}^{-3})$ terms, integrate them out, and
lastly stabilize $\{\Phi_{III}\}$ by the remaining potential.

\section{Toward Modular Inflation \label{section:app}}
Understanding the moduli stabilization problem using the language of
effective theory, one can guess that the following will be valid
approaches to simplify inflation analysis:
\begin{enumerate}
\item Integrating out the complex structure moduli and the axion-dilaton
and then using the remaining theory to find inflation.
\item Integrating out the complex structure moduli, axion-dilaton, $b_s$, $\tau_s$, and
$\cal V$ and then using the remaining theory to find inflation.
\end{enumerate}
We will see that these approaches are valid by analyzing the
equations of motion from the full theory. The necessary metric,
inverse metric, and connection for the following subsections are
given in Appendix \ref{section:tool}.

\subsection{First Approach}
Basically, we are trying to decouple complex structure moduli and
the axion-dilaton from inflationary dynamics. First, we turn on
\emph{only} the fluxes needed to stabilize the complex structure moduli and
the axion-dilaton, so that $\Phi_I = {\Phi_{I}}_0$. Then, we
incorporate the non-perturbative effects to create a potential for
the K\"{a}hler moduli, which will be the inflatons.

Let $\Phi_I = {\Phi_{I}}_0 + \chi$, where ${\Phi_{I}}_0 >> \chi$ and
$\dot{\chi}(t = 0) = 0$. Let us also assume that the inflatons are
in the inflationary regime. Then, the fluctuations of the complex structure moduli about their minima satisfy the following equation at $t = 0$:
\begin{equation}
\ddot{\chi}^a + 2 G^{a b} V_{,b} + 2 G^{a \tau} V_{,\tau}= 0 .
\end{equation}
Since ${\Phi_{I}}_0$ is at the minimum of the leading terms of the
potential $\ddot{\chi}^a \sim {\cal O}({\cal V}^{-3})$.

Similarly, for the axion-dilaton, at $t = 0$,
\begin{equation}
{\ddot{\chi}}^{\tau} + {\Gamma^{\tau}}_{\tau_l \tau_m} \dot{\phi}^l \dot{\phi}^m
+ {\Gamma^{\tau}}_{\tau_l \tau_s} \dot{\phi}^l \dot{\phi}^s +
{\Gamma^{\tau}}_{\tau_s \tau_s} \dot{\phi}^s \dot{\phi}^s + 2 \left( G^{\tau
\bar{\tau}} V_{,\bar{\tau}} + G^{\tau a} V_{,a} + G^{\tau \tau_l}
V_{,\tau_l} + G^{\tau \tau_s} V_{,\tau_s} \right)= 0,
\end{equation}
where $\phi^i$ can be either the axion $b_i$ or the 4-cycle modulus
$\tau_i$. Again, since ${\Phi_{I}}_0$ is at the minimum of the
leading terms of the potential, $G^{\tau \bar{\tau}} V_{,\bar{\tau}}
\sim G^{\tau a} V_{,a} \sim {\cal O}({\cal V}^{-3})$, while
$G^{\tau \tau_l} V_{,\tau_l} \sim {\cal O}({\cal V}^{-13\slash 3})$ and $G^{\tau \tau_s} V_{,\tau_s} \sim {\cal O}({\cal V}^{-4})$. Furthermore, since we are assuming that
slow-roll obtains,
\begin{eqnarray}
{\Gamma^{\tau}}_{\tau_l \tau_m} \dot{\phi}^l \dot{\phi}^m &\lesssim&
\frac{1}{{\cal V}^{7/3}} \frac{V}{G_{\tau_l \tau_m}} \sim
\frac{1}{{\cal V}^4}, \nonumber \\
{\Gamma^{\tau}}_{\tau_l \tau_s} \dot{\phi}^l \dot{\phi}^s &\lesssim&
\frac{1}{{\cal V}^{8/3}} \frac{V}{G_{\tau_l \tau_s}} \sim \frac{1}{{\cal V}^4}, \nonumber \\
{\Gamma^{\tau}}_{\tau_s \tau_s} \dot{\phi}^s \dot{\phi}^s &\lesssim&
\frac{1}{{\cal V}^2} \frac{V}{G_{\tau_s \tau_s}} \sim \frac{1}{{\cal
V}^4}.
\end{eqnarray}
Putting all these results together tells us that ${\ddot{\chi}}^{\tau}(t=0) \sim {\cal O}({\cal V}^{-3})$.

At the next instant $\Delta t > 0$, $\dot{\chi}(\Delta t) =
\dot{\chi} (0) + \ddot{\chi} \Delta t = \ddot{\chi} \Delta t$. This
Taylor's expansion is valid only for small $\Delta t$, and since
$H^{-1} \sim {\cal V}^{3/2}$, the only small time scale in the
theory is the string scale, which is equal to 1. Therefore,
$\dot{\chi}^a \sim \dot{\chi}^{\tau} \sim {\cal O}({\cal V}^{-3})$.
The terms $3 H \dot{\chi} \sim {\cal O}({\cal V}^{-9\slash 2})$,
$\Gamma \, \dot{\chi} \dot{\chi} \sim {\cal O}({\cal V}^{-6})$, and
$\Gamma \, \dot{\chi} \dot{\phi} \sim {\cal O}({\cal V}^{-14\slash
3})$ still cannot compete with the derivative of the potential.
Thus, as long as we are in the inflationary regime, ${\ddot{\chi}}
\sim {\cal O}({\cal V}^{-3})$ and $\dot{\chi} \sim {\cal O}({\cal
V}^{-3})$. Therefore, the contribution of the complex structure
moduli and the axion-dilaton to the slow-roll parameter $\epsilon$
is
\begin{equation}
\epsilon_{cs} \sim \frac{G \, {\dot{\chi}} {\dot{\chi}}}{V} \sim
\frac{1}{{\cal V}^3}.
\end{equation}
Thus, as long as $\frac{G \, {\dot{\phi}} {\dot{\phi}}}{V} < 1$, we
do not have to worry about the contribution from $\{\Phi_{I}\}$.
Therefore, we can decouple them from inflation analysis.

Furthermore, calculating order of magnitudes, one should be able to
see that the contributions of $\chi$ in the equation of motion of
the inflatons are in fact negligible, thus validating equation
(\ref{equation:eom_of_inflatons}).

An example of inflationary model where only the complex structure
moduli and the axion-dilaton are stabilized can be found in
\cite{racetrack}\footnote{Even though \cite{racetrack} does not
include the $\alpha'$ corrections,  it should be possible to extend their analysis to include them.}.

\subsection{Second Approach}
In this approach, after fixing $\{{\Phi_{I}}\}$, instead of turning
on all the non-perturbative effects at once, we turn on only the one
that corresponds to $\tau_s$, so that $\Phi_I = {\Phi_{I}}_0 + {\cal
O}(\frac{1}{\cal V})$ and $\Phi_{II} = {\Phi_{II}}_0$. Next, we turn
on the rest of the non-perturbative effects to switch on the
potential for $\{{\Phi_{III}}\}$. Let $\Phi_I = {\Phi_{I}}_0 +
\chi_I$, where $\chi_I << \frac{1}{\cal V}$ and $\dot{\chi}_I(t = 0)
= 0$, and let $\Phi_{II} = {\Phi_{II}}_0 + \chi_{II}$, where
$\Phi_{II} >> \chi_{II}$ and $\dot{\chi}_{II}(t = 0) = 0$. Let us
also assume that $\{{\Phi_{III}}\}$ is in the inflationary regime.

Note that the potential is exponentially suppressed. On the other hand,  all
geometry related quantities are $\sim {\cal V}^{\alpha}$ and
even if $\alpha > 0$, the geometrical quantities cannot compete with
$e^{{\cal V}^{2/3}}$ in the denominator. Because all we want to say is that
the contribution from the kinetic energy of $\{{\Phi_{I}}\}$ and
$\{{\Phi_{II}}\}$ terms toward $\epsilon$ is negligible, we can neglect
the ${\cal V}^{\alpha}$ factor and concentrate only on the
exponential factor.

This allows us to infer that, ${\ddot{\chi}}_{I, II}(t = 0) \sim e^{- {\cal V}^{2/3}}$. Following the analysis done above for the first approach, we can say that a string time
later ${\dot{\chi}}_{I, II} \sim e^{- {\cal V}^{2/3}}$ so that $3 H
\dot{\chi} \sim  e^{- 2{\cal V}^{2/3}}$ and $\Gamma \, \dot{\chi}
\dot{\chi} \sim e^{- 2{\cal V}^{2/3}}$, which means that these terms are negligible
compared to the other terms in the equation of motion. Thus, as long
as $\{{\Phi_{III}}\}$ are in the inflationary regime, ${\ddot{\chi}}
\sim e^{- {\cal V}^{2/3}}$ and $\dot{\chi} \sim e^{- {\cal V}^{2/3}}$.
Therefore, the contribution of $\{{\Phi_{I}}\}$ and
$\{{\Phi_{II}}\}$ to $\epsilon$ is
\begin{equation}
\epsilon_{{\Phi}_{I, II}} \sim \frac{G \, {\dot{\chi}}
{\dot{\chi}}}{V} \sim \frac{1}{e^{{\cal V}^{2/3}}}.
\end{equation}
Therefore, we can decouple $\{{\Phi_{I}}\}$ and $\{{\Phi_{II}}\}$
from inflation analysis, and only worry about finding inflationary
regime for $\{{\Phi_{III}}\}$.

An example where one can get a single-field inflation from this
approach is given in \cite{axinfle}.

\section{Oscillation Effects on the Spectrum \label{section:spectrum}}

We have seen that the large volume limit allows us to decouple a sufficient number of the moduli so that the problem of finding inflationary phases becomes tractable. However, when we say that the stabilized moduli are at the potential minimum, we mean that the {\em zero mode} is frozen. The fluctuations  around this zero mode could be oscillating about the minimum and this  may give rise to interesting effects~\cite{transplanckianme}. In particular, these oscillations could imprint themselves on the CMB power spectrum; it should be noted that the exact nature of the effect depends on the model.

We need to be mindful of the requirement that the amplitude of oscillations be small enough that the energy density contained in them not disrupt the inflationary phase. This can be accomplished by just waiting long enough, since the energy density in these oscillations decays as that of non-relativistic matter.

What we would like to do is to calculate the power spectrum of the inflatons in the presence of these oscillations of the complex-structure moduli. To do this completely is a difficult problem, but we can at least estimate the order of magnitude of the effect within the large volume approximation scheme used above. We will see that the two approaches dealt with above can give rise to very different, and potentially measurable results.

\subsection{First Approach}
Let us consider the quantum fluctuations of the inflatons, $\delta
{{\phi}^l}$. A mode with wave number $\bf k$ has the following
equation of motion
\begin{eqnarray}
\delta {{\ddot{\phi}}^l}_{\bf k} &+& 3 H \delta
{{\dot{\phi}}^l}_{\bf k} + {\Gamma^{\tau_l}}_{t u} \left(\,
\dot{\phi}^t \delta {{\dot{\phi}}^u}_{\bf k} + \delta
{{\dot{\phi}}^t}_{\bf k} \dot{\phi}^u \, \right) +
{\Gamma^{\tau_l}}_{t u, v} {\dot{\phi}}^t \dot{\phi}^u \delta
{\phi}^v_{\bf k} + \left(G^{\tau_l t} V_{, t u} + {G^{\tau_l
t}}_{,u} V_{, t} \right)\delta
{\phi}^u_{\bf k} \nonumber \\
&+& \left|k\right|^2 e^{-2Ht} \delta {\phi}^l_{\bf k} = 2 \,
{\Gamma^{\tau_l}}_{\tau t} {\dot{\chi}}^{\tau} \delta
{{\dot{\phi}}^t}_{\bf k} + 2 \, {\Gamma^{\tau_l}}_{\tau t , u}
{\dot{\chi}}^{\tau} {\dot{\phi}}^t \delta {\phi}^u_{\bf k} \nonumber \\
&+& \left(
{\Gamma^{\tau_l}}_{\tau \bar{\tau} , t} {\dot{\chi}}^{\tau}
{\dot{\chi}}^{\tau} + G^{\tau_l \tau} V_{, \tau t} + {G^{\tau_l
\tau}}_{, t} V_{, \tau} \right) \delta {\phi}^t_{\bf k},
\label{eq:quantum_large}
\end{eqnarray}
where $t$, $u$, and $v$ can be $\tau_s$ or $\tau_m$ with $m \ne s$. If there were no oscillating $\chi$ fields, the equation for the fluctuations would be the one above, but with the right hand side set to zero. Now let's turn to estimating the order of magnitude of the various terms in the Eq.~(\ref{eq:quantum_large}):
\begin{eqnarray}
{\rm LHS} &\sim& \delta {{\ddot{\phi}}^l}_{\bf k} + \left( {\cal
O}(\frac{1}{{\cal V}^{4/3}}) \, \delta {{\dot{\phi}}^s}_{\bf k} +
{\cal O}(\frac{1}{{\cal V}^{7/3}}) \delta {\phi}^s_{\bf k} \right) +
\sum_{m \ne s} \left( {\cal O}(\frac{1}{{\cal V}^{3/2}}) \, \delta
{{\dot{\phi}}^m}_{\bf k} + {\cal
O}(\frac{1}{{\cal V}^3}) \, \delta {\phi}^m_{\bf k}\right) \nonumber \\
&+& \left|k\right|^2 e^{-2Ht} \delta {\phi}^l_{\bf k}, \nonumber\\
{\rm RHS} &\sim& \left( {\cal O}(\frac{1}{{\cal V}^{13/3}}) \,
\delta {{\dot{\phi}}^s}_{\bf k} + {\cal O}(\frac{1}{{\cal
V}^{11/3}}) \delta {\phi}^s_{\bf k} \right)
+ \sum_{m \ne s} \left(
{\cal O}(\frac{1}{{\cal V}^4}) \, \delta {{\dot{\phi}}^m}_{\bf k} +
{\cal O}(\frac{1}{{\cal V}^{13/3}}) \, \delta {\phi}^m_{\bf
k}\right).
\end{eqnarray}

Similarly, for the mode $\delta {\phi}^s_{\bf k}$, we get
\begin{eqnarray}
\delta {{\ddot{\phi}}^s}_{\bf k} &+& 3 H \delta
{{\dot{\phi}}^s}_{\bf k} + {\Gamma^{\tau_s}}_{t u} \left(\,
\dot{\phi}^t \delta {{\dot{\phi}}^u}_{\bf k} + \delta
{{\dot{\phi}}^t}_{\bf k} \dot{\phi}^u \, \right) +
{\Gamma^{\tau_s}}_{t u, v} {\dot{\phi}}^t \dot{\phi}^u \delta
{\phi}^v_{\bf k} + \left(G^{\tau_s t} V_{, t u} + {G^{\tau_s
t}}_{,u} V_{, t}\right) \delta
{\phi}^u_{\bf k} \nonumber \\
&+& \left|k\right|^2 e^{-2Ht} \delta {\phi}^s_{\bf k} = 2 \,
{\Gamma^{\tau_s}}_{\tau t} {\dot{\chi}}^{\tau} \delta
{{\dot{\phi}}^t}_{\bf k} + 2 {\Gamma^{\tau_s}}_{\tau t , u}
{\dot{\chi}}^{\tau} {\dot{\phi}}^t \delta {\phi}^u_{\bf k} \nonumber\\
& +& \left(
{\Gamma^{\tau_s}}_{\tau \bar{\tau} , t} {\dot{\chi}}^{\tau}
{\dot{\chi}}^{\tau} + G^{\tau_s \tau} V_{, \tau t} + {G^{\tau_s
\tau}}_{, t} V_{, \tau} \right) \delta {\phi}^t_{\bf k}.
\label{eq:quantum_small}
\end{eqnarray}
Examining the left and right hand sides of this equation gives us:
\begin{eqnarray}
{\rm LHS} &\sim& \delta {{\ddot{\phi}}^s}_{\bf k} + \left( {\cal
O}(\frac{1}{{\cal V}^{3/2}}) \, \delta {{\dot{\phi}}^s}_{\bf k} +
{\cal O}(\frac{1}{{\cal V}^2}) \delta {\phi}^s_{\bf k} \right) +
\sum_{m \ne s} \left( {\cal O}(\frac{1}{{\cal V}^{5/3}}) \, \delta
{{\dot{\phi}}^m}_{\bf k} + {\cal O}(\frac{1}{{\cal V}^{8/3}}) \,
\delta {\phi}^m_{\bf k}\right) \nonumber\\
&+& \left|k\right|^2 e^{-2Ht}
\delta {\phi}^s_{\bf k}, \nonumber \\
{\rm RHS} &\sim& {\cal O}(\frac{1}{{\cal V}^4}) \left(\, \delta
{{\dot{\phi}}^s}_{\bf k} + \delta {\phi}^s_{\bf k} \right) + \sum_{m
\ne s} {\cal O}(\frac{1}{{\cal V}^{14/3}}) \, \left( \delta
{{\dot{\phi}}^m}_{\bf k} + \delta {\phi}^m_{\bf k} \right).
\end{eqnarray}

The equations of motion for the modes then become
\begin{eqnarray}
\delta {{\ddot{\phi}}^l}_{\bf k} &+& \left|k\right|^2 e^{-2Ht}
\delta {\phi}^l_{\bf k} + \left[ {\cal O}(\frac{1}{{\cal V}^{4/3}})
\, \left( 1 + {\cal O}(\frac{1}{{\cal V}^3}) \right) \, \delta
{{\dot{\phi}}^s}_{\bf k} + {\cal O}(\frac{1}{{\cal V}^{7/3}}) \left(
1 + {\cal O}(\frac{1}{{\cal
V}^{4/3}}) \right) \, \delta {\phi}^s_{\bf k} \right] \nonumber \\
&+& \sum_{m \ne s} \left[ {\cal O}(\frac{1}{{\cal V}^{3/2}}) \,
\left( 1 + {\cal O}(\frac{1}{{\cal V}^{5/2}}) \right) \, \delta
{{\dot{\phi}}^m}_{\bf k} + {\cal O}(\frac{1}{{\cal V}^3}) \, \left(
1 + {\cal O}(\frac{1}{{\cal V}^{4/3}}) \right) \, \delta
{\phi}^m_{\bf k}\right] = 0, \\
\delta {{\ddot{\phi}}^s}_{\bf k} &+& \left|k\right|^2 e^{-2Ht}
\delta {\phi}^s_{\bf k} + \left[ {\cal O}(\frac{1}{{\cal V}^{3/2}})
\, \left( 1 + {\cal O}(\frac{1}{{\cal V}^{5/2}}) \right) \, \delta
{{\dot{\phi}}^s}_{\bf k} + {\cal O}(\frac{1}{{\cal V}^2}) \left( 1 +
{\cal O}(\frac{1}{{\cal V}^2}) \right) \, \delta
{\phi}^s_{\bf k} \right] \nonumber \\
&+& \sum_{m \ne s} \left[ {\cal O}(\frac{1}{{\cal V}^{5/3}}) \,
\left( 1 + {\cal O}(\frac{1}{{\cal V}^{9/3}}) \right) \, \delta
{{\dot{\phi}}^m}_{\bf k} + {\cal O}(\frac{1}{{\cal V}^{8/3}}) \,
\left( 1 + {\cal O}(\frac{1}{{\cal V}^2}) \right) \, \delta
{\phi}^m_{\bf k}\right] = 0,
\end{eqnarray}
and we can neglect the contributions from $\chi$. Therefore, in
order to calculate the spectrum, we only need to solve
\begin{equation}
{\rm LHS \, \, of \, \, eq. \, \, (\ref{eq:quantum_large})} = {\rm
LHS \, \, of \, \, eq. \, \, (\ref{eq:quantum_small})} = 0.
\end{equation}

Furthermore, we can estimate the effect of the oscillation of
$\{\chi\}$ in the spectrum. Since all the coefficients in front of
$\delta {\phi}_{\bf k}$ and $\delta {\dot{\phi}}_{\bf k}$ are in the
form of $A (1 + B)$ with $B << 1$, the ratio of the effect of
$\{\chi\}$ in the spectrum with the spectrum will be the biggest
$B$. Thus,
\begin{equation}
\frac{\delta P}{P}(\left|{\bf k}\right|) \sim \frac{1}{{\cal
V}^{4/3}}.
\end{equation}
For the model in \cite{racetrack}, the volume is ${\cal V} = 99$ in
string units. Thus, the change in the spectrum from
complex-structure moduli and the axion-dilaton is of order
$10^{-2}$. Since the current experiment can measure $\delta P \slash
P$ up to order $10^{-3} - 10^{-2}$, it is necessary to calculate
this effect in model \cite{racetrack}\footnote{To do so, one has to
extend the analysis in \cite{racetrack} to include the $\alpha'$
corrections.}.

\subsection{Second Approach}
For the second approach, since $\dot{\chi} \sim e^{-{\cal V}^{2/3}}$ and
$V \sim e^{-{\cal V}^{2/3}}$ then in the equation of motion, the
contribution of $\chi$ will also be of order $e^{-{\cal V}^{2/3}}$. On the
other hand, the other terms are of order $\sqrt{V} \sim e^{-{\cal
V}^{2/3}/2}$. Thus, the effect of $\chi$'s oscillation on the spectrum is
\begin{equation}
\frac{\delta P}{P}(\left|{\bf k}\right|) \sim \frac{1}{e^{{\cal
V}^{2/3}/2}}.
\end{equation}
Since the volume is at least $10^2 - 10^3$ string units, this effect is too small to be measured.

\section{Discussion}
Motivated by the scale hierarchy of the moduli in the large volume
scheme, we have approached the problem of moduli stabilization by
dividing it into several stages. We would like to emphasize that the
decoupling in the moduli stabilization procedure does {\em not} come
from any underlying assumptions such as suggested in the original
KKLT procedure \cite{kklt}. The decoupling comes from approaching
this problem perturbatively using $1/{\cal V}$-expansion.
Furthermore, requiring the $\alpha'$-expansion to be valid leads us to $R > \sqrt{\alpha'}$, where $R$ is the
'average' radius of the compact dimensions. Thus, working at the
large volume regime is a natural thing to do and with that, the
decoupling in our three-step moduli stabilization comes naturally
too. Therefore, modifying KKLT procedure as suggested by
\cite{modkklt} might not be necessary.

We also have shown that the fields that are stabilized in the
earlier stage(s) can be integrated out of the theory, thus reducing
the number of possible inflatons and rendering the search for an inflationary phase in this theory easier
\cite{axinfle}.

While we did not pursue the detailed analysis of this possibility,
we also have seen that the oscillation of the stabilized fields
could, at least in principle, modify the scalar power spectrum. For
the second method, this modification is small and cannot be measured
by our current experiments. Thus, we can calculate the power
spectrum as if there is no oscillating fields in the background.
However, we saw that in the first approach, there is the possibility
that an effect could be observable. This merits further study.

In this paper, we have assumed the use of non-perturbative effects
from D3-instantons or gaugino condensations with low rank gauge
group (i.e.: small $N$, $a_i$ of ${\cal O}(1)$). However, there are
many models where $N$ needs to be large. For moderate $N$, as long
as all $a_i$'s are of the same order of magnitude, the hierarchy we
have described still exist, only with smaller gaps between the
stages. Therefore, most of our arguments here are applicable to the
cases with larger $N$, with the exception that there is a
possibility that the modification of the power spectrum in the
second method can be larger and thus, observable. If $N$ gets to a
comparable size as the stabilized volume, then our expansion is no longer valid. Furthermore, at large $N$, higher instantons corrections must be included in the superpotential.   

As noted in \cite{vijay}, our arguments may not be completely
airtight. The treatment of the the loop determinant $A_i$ as a
constant, may not be warranted. In particular, if $A_i$ depends on
the K\"{a}hler moduli, our argument might not be valid. If $A_s \sim
{\cal V}^{\alpha}$ (we do not have to worry for $A_l$ due to the
exponential-suppression on the denominator), we can save our
argument by redefining $\tau_s \sim (\alpha + 1) \ln {\cal V}$.
However, the polynomial dependence on the K\"{a}hler moduli is
unlikely due to holomorphy and shift symmetry.

In the previous sections, we have deliberately used the language of
"turning on" fluxes and non-perturbative effects in a certain order.
This has to be understood as a mathematical tool to simplify the
calculation. We are not suggesting that nature has to do so in order
for the moduli to be stabilized. Even when both fluxes and
non-perturbative effects are switched on at the same time,
$\{{\Phi_{II}}\}$ will roam around until $\{{\Phi_{I}}\}$ are
stabilized, and after that, they will roll toward the minimum.
Similarly, $\{{\Phi_{III}}\}$ will wait until both $\{{\Phi_{I}}\}$
and $\{{\Phi_{II}}\}$ are stabilized before rolling toward the
minimum. Nevertheless, it is \emph{not} impossible that nature chose
to turn different fluxes and non-perturbative effects at different
times. Whether that was the case in the evolution of our universe
remains an open question. Answering this question requires a deeper
understanding in time-dependent background in string theory in
particular and background independence in general.

From the point of view of inflationary dynamics, there is also an issue
of the likelihood of the initial conditions. Given that
$\{{\Phi_{I}}\}$ for the first approach (or $\{{\Phi_{I}}\}$ and
$\{{\Phi_{II}}\}$ for the second approach) are at the minimum, how
likely will it be for the rest of the moduli to be in the slow-roll
regime? This requires further analysis.

We also would like to emphasize that our approaches might not be the
only way to simplify the analysis of inflation in flux compactifications. A
different approach would be to change the definition of the large volume
limit. Nevertheless, the trick will be the same, namely exploitation
of the scale hierarchy of the moduli. Thus, 'decoupling' a field $\psi$
from the inflationary dynamics by 'constraining' it to the minima,
while letting inflaton $\phi$ rolls over a potential $V(\phi)$ that
has a comparable scale to the potential for $\psi$ will not be
valid.

In the literature, there are inflationary models where all axions
are integrated out (e.g.: \cite{kahler, topology}). This is somewhat counter-intuitive from the point of view of effective field theory approach, since we are integrating out some of the lighter fields. Nevertheless, close inspection of the equation of motion shows that this method is in fact valid\footnote{We thank V. Balasubramanian for discussion about this point.}.

One possible alternative definition of the large volume limit is the limit where only one 4-cycle modulus gets large, $\tau_l \rightarrow \infty$, while the rest of the 4-cycle moduli are finite, $\tau_{i \ne l} \sim \ln {\cal V}$. This will result in having single-field inflationary models as in \cite{axinfle} without having the extra restriction $h^{1,1} = 2$ as required in that reference.

It would also be interesting to see whether there is a correlation
between the number of left-over moduli and the power spectrum. If
there is, then as cosmological data becomes more precise, it would not be
surprising that one can put constraints on the extra dimensions
using cosmological data (an initial attempt at falsifying stringy inflationary models was given
in Ref.~\cite{topology}).

\section{Acknowledgment}
This work was supported in part by DOE grant DE-FG03-91-ER40682.

\appendix
\section{Slow-roll Condition for Multi-Field Inflation \label{section:slow-roll}}
Consider an FRW background. From Einstein equations, we can get the
evolution of the scale factor
\begin{equation}
H^2 \equiv {\left( \frac{\dot{a}}{a} \right)}^2 = \frac{\rho}{6},
\label{eq:friedmann}
\end{equation}
where assuming homogeneity and isotropy, the energy density of the
system $\rho = G_{AB} \dot{\phi}^{A} \dot{\phi}^{B} + V$. Using
Friedmann equation (\ref{eq:friedmann}) and the mass conservation,
we get the equation for the acceleration of the scale factor
\begin{equation}
\frac{\ddot{a}}{a} = - \frac{\rho + 3 p}{6}, \label{eq:acc}
\end{equation}
where $p = G_{AB} \dot{\phi}^{A} \dot{\phi}^{B} - V$.

Inflation is defined as an epoch where $\ddot{a} \slash a > 0$.
Since
\begin{equation}
\frac{\ddot{a}}{a} = H^2 (1 - \epsilon); \, \, \, \epsilon \equiv -
\, \frac{\dot{H}}{H^2} = \frac{G_{AB} \dot{\phi}^{A}
\dot{\phi}^{B}}{H^{2}},
\end{equation}
inflation $\Leftrightarrow \epsilon < 1$. Notice that from equation
(\ref{eq:acc}), inflation also means that
\begin{equation}
2 G_{AB} \dot{\phi}^{A} \dot{\phi}^{B} < V. \label{eq:cond1}
\end{equation}
If we further assume that
\begin{equation}
2 G^{AB} V_{,B} >> \ddot{\Phi}^{A} + {\Gamma^{A}}_{BC}
\dot{\Phi}^{B} \dot{\Phi}^{C}, \label{eq:cond2}
\end{equation}
we get
\begin{equation}
\epsilon = \frac{G^{AB} V_{,A} V_{,B}}{4 V^2}.
\end{equation}
Up until this point, this analysis resembles the one for the case of
single-field inflation. In single-field inflation, one will get
another condition $\eta < 1$ for inflation by demanding equation
(\ref{eq:cond1}) is consistent with equation (\ref{eq:cond2}).
However, in multi-field inflation, one is not able to define $\eta$
in that manner. Since our main discussion does not involve $\eta$,
we will not discuss this matter any further\footnote{One possibility
in defining $\eta$ is given in \cite{realistic}.}.

\section{The Dependence of Metric, Inverse Metric, and Connection on Classical Volume \label{section:tool}}

In the large volume limit, the K\"{a}hler potential becomes
\begin{eqnarray}
K &=& 3 \phi_0 - \log \left[-i \int_{M} \Omega \wedge \bar{\Omega}
\right] - \log \left[-i (\tau - \bar{\tau}) \right] - 2 \log
\left[\,1 + \frac{e^{3 \phi_{0} /2}}{\cal V} \frac{\xi}{2} \left(
\frac{- i (\tau - \bar{\tau})}{2} \right)^{3/2}\, \right] \nonumber
\\
& & - 2 \log
{\cal V}, \nonumber \\
&=& 3 \phi_0 - \log \left[-i \int_{M} \Omega \wedge \bar{\Omega}
\right] - \log \left[-i (\tau - \bar{\tau}) \right] - \frac{2 e^{3
\phi_{0} /2}}{\cal V} \frac{\xi}{2} \left( \frac{- i (\tau -
\bar{\tau})}{2} \right)^{3/2} \nonumber \\
& & - 2 \log {\cal V}.
\end{eqnarray}
Noticing that the relation between volume and the four-cycle moduli
is like ${\cal V} \sim {\tau_i} t^i$, and that ${\cal V} \sim
{\tau_l}^{3/2}$, $l \ne s$ in the large volume limit, we get
\begin{equation}
\frac{\partial {\cal V}}{\partial \tau_l} \sim t^l \sim
{\tau_l}^{1/2} \sim {\cal V}^{1/3},
\end{equation}
for $l \ne s$, and
\begin{equation}
\frac{\partial {\cal V}}{\partial \tau_s} \sim t^s \sim
{\tau_s}^{1/2} \sim {\cal O}(1) .
\end{equation}
Therefore, the components of the metric become
\begin{eqnarray}
G_{\tau \bar{\tau}} &=& {\cal O}(1), \,\,\,\,\, G_{\tau
\bar{\rho_l}} \sim
\frac{1}{{\cal V}^{5/3}}, \,\,\,\,\, G_{\tau \bar{\rho_s}} \sim \frac{1}{{\cal V}^2}, \nonumber \\
G_{\rho_l \bar{\rho_m}} &\sim& \frac{1}{{\cal V}^{4/3}}, \,\,\,\,\,
G_{\rho_l \bar{\rho_s}} \sim \frac{1}{{\cal V}^{5/3}}, \,\,\,
G_{\rho_s \bar{\rho_s}} \sim \frac{1}{\cal V}.
\end{eqnarray}
The components of the inverse metric are given in \cite{bobkov}.
\begin{eqnarray}
G^{\tau \bar{\tau}} &=& {\cal O}(1), \,\,\,\,\, G^{\tau
\bar{\rho_l}} \sim
\frac{1}{{\cal V}^{2/3}}, \,\,\,\,\, G^{\tau \bar{\rho_s}} \sim \frac{1}{\cal V}, \nonumber \\
G^{\rho_l \bar{\rho_m}} &\sim& {\cal V}^{4/3}, \,\,\,\,\, G^{\rho_l
\bar{\rho_s}} \sim {\cal V}^{2/3}, \,\,\,\,\, G^{\rho_s
\bar{\rho_s}} \sim {\cal V}.
\end{eqnarray}

Let us remind ourselves that we need to change variables from the
complex moduli fields to the real scalar fields for calculation in
Section \ref{section:app}. Since the components of the metric (and
inverse metric) for the real scalar fields are of the same order
with the corresponding components for the complex moduli fields, we
will adopt a somewhat loose notation for the connection. The
components of the connection necessary for calculation in Section
\ref{section:app} and Section \ref{section:spectrum} are
\begin{eqnarray}
{\Gamma^{\tau}}_{\tau_l \tau_m} &=& \frac{1}{2} \, G^{\tau
\bar{\tau}} \left(\, G_{\tau_l \bar{\tau} , \tau_m} + G_{\tau_m
\bar{\tau} , \tau_l} -
G_{\tau_l \tau_m, \bar{\tau}} \, \right) \nonumber \\
& & + \frac{1}{2} \, G^{\tau \tau_n} \left(\, G_{\tau_l \tau_n ,
\tau_m} + G_{\tau_m \tau_n , \tau_l} - G_{\tau_l \tau_m, \tau_n} \,
\right)
\nonumber \\
& & + \frac{1}{2} \,  G^{\tau \tau_s} \left(\, G_{\tau_l \tau_s ,
\tau_m} + G_{\tau_m \tau_s , \tau_l} - G_{\tau_l \tau_m, \tau_s} \,
\right),
\nonumber \\
&\sim& {\cal O}(1) \, \frac{1}{{\cal V}^{7/3}} +  \frac{1}{{\cal
V}^{2/3}} \, \frac{1}{{\cal V}^2} + \frac{1}{\cal V} \,
\frac{1}{{\cal V}^{7/3}}, \nonumber \\
&\sim& \frac{1}{{\cal V}^{7/3}}, \\
{\Gamma^{\tau}}_{\tau_l \tau_s} &=& \frac{1}{2} \, G^{\tau
\bar{\tau}} \left(\, G_{\tau_l \bar{\tau} , \tau_s} + G_{\tau_s
\bar{\tau} , \tau_l} -
G_{\tau_l \tau_s, \bar{\tau}} \, \right) \nonumber \\
& & + \frac{1}{2} \,  G^{\tau \tau_m} \left(\, G_{\tau_l \tau_m ,
\tau_s} +
G_{\tau_s \tau_m , \tau_l} - G_{\tau_l \tau_s, \tau_m} \, \right) + \frac{1}{2} \,  G^{\tau \tau_s} \, G_{\tau_s \tau_s , \tau_l}, \nonumber \\
&\sim& {\cal O}(1) \, \frac{1}{{\cal V}^{8/3}} +  \frac{1}{{\cal
V}^{2/3}} \, \frac{1}{{\cal V}^{7/3}} + \frac{1}{\cal V} \,
\frac{1}{{\cal V}^{5/3}}, \nonumber \\
&\sim& \frac{1}{{\cal V}^{8/3}}, \\
{\Gamma^{\tau}}_{\tau_s \tau_s} &=& \frac{1}{2} \, G^{\tau
\bar{\tau}} \left(\, 2 G_{\tau_s \bar{\tau} , \tau_s} -
G_{\tau_s \tau_s, \bar{\tau}} \, \right) \nonumber \\
& & + \frac{1}{2} \,  G^{\tau \tau_l} \left(\, 2 G_{\tau_s \tau_l ,
\tau_s} - G_{\tau_s \tau_s, \tau_l} \, \right) + \frac{1}{2} \,
G^{\tau \tau_s} \, G_{\tau_s \tau_s, \tau_s},
\nonumber \\
&\sim& {\cal O}(1) \, \frac{1}{{\cal V}^2} +  \frac{1}{{\cal
V}^{2/3}} \, \frac{1}{{\cal V}^{5/3}} + \frac{1}{\cal V} \,
\frac{1}{\cal V}, \nonumber \\
&\sim& \frac{1}{{\cal V}^2}, \\ 
{\Gamma^{\tau_s}}_{\tau_l \tau_m} &=& \frac{1}{2} \, G^{\tau_s
\bar{\tau}} \left(\, G_{\tau_l \bar{\tau} , \tau_m} + G_{\tau_m
\bar{\tau} , \tau_l} -
G_{\tau_l \tau_m, \bar{\tau}} \, \right) \nonumber \\
& & + \frac{1}{2} \,  G^{\tau_s \tau_n} \left(\, G_{\tau_l \tau_n ,
\tau_m} + G_{\tau_m \tau_n , \tau_l} - G_{\tau_l \tau_m, \tau_n} \,
\right)
\nonumber \\
& & + \frac{1}{2} \,  G^{\tau_s \tau_s} \left(\, G_{\tau_l \tau_s ,
\tau_m} + G_{\tau_m \tau_s , \tau_l} - G_{\tau_l \tau_m, \tau_s} \,
\right),
\nonumber \\
&\sim& \frac{1}{\cal V} \, \frac{1}{{\cal V}^{7/3}} + {\cal V}^{2/3}
\, \frac{1}{{\cal V}^2} + {\cal V} \,
\frac{1}{{\cal V}^{7/3}}, \nonumber \\
&\sim& \frac{1}{{\cal V}^{4/3}}, \\
{\Gamma^{\tau_s}}_{\tau_l \tau_s} &=& \frac{1}{2} \, G^{\tau_s
\bar{\tau}} \left(\, G_{\tau_l \bar{\tau} , \tau_s} + G_{\tau_s
\bar{\tau} , \tau_l} -
G_{\tau_l \tau_s, \bar{\tau}} \, \right) \nonumber \\
& & + \frac{1}{2} \,  G^{\tau_s \tau_m} \left(\, G_{\tau_l \tau_m ,
\tau_s} +
G_{\tau_s \tau_m , \tau_l} - G_{\tau_l \tau_s, \tau_m} \, \right) + \frac{1}{2} \,  G^{\tau_s \tau_s} \, G_{\tau_s \tau_s , \tau_l}, \nonumber \\
&\sim& \frac{1}{\cal V} \, \frac{1}{{\cal V}^{8/3}} + {\cal V}^{2/3}
\, \frac{1}{{\cal V}^{7/3}} + {\cal V} \,
\frac{1}{{\cal V}^{5/3}}, \nonumber \\
&\sim& \frac{1}{{\cal V}^{2/3}}, \\
{\Gamma^{\tau_s}}_{\tau_s \tau_s} &=& \frac{1}{2} \, G^{\tau_s
\bar{\tau}} \left(\, 2 G_{\tau_s \bar{\tau} , \tau_s} -
G_{\tau_s \tau_s, \bar{\tau}} \, \right) \nonumber \\
& & + \frac{1}{2} \,  G^{\tau_s \tau_l} \left(\, 2 G_{\tau_s \tau_l
, \tau_s} - G_{\tau_s \tau_s, \tau_l} \, \right) + \frac{1}{2} \,
G^{\tau_s \tau_s} \, G_{\tau_s \tau_s, \tau_s},
\nonumber \\
&\sim& \frac{1}{\cal V} \, \frac{1}{{\cal V}^2} +  {\cal V}^{2/3} \,
\frac{1}{{\cal V}^{5/3}} + {\cal V} \,
\frac{1}{\cal V}, \nonumber \\
&\sim& \frac{1}{{\cal V}^2}, \\
{\Gamma^{\tau_s}}_{\tau \tau_l} &=& \frac{1}{2} \, G^{\tau_s
\bar{\tau}} \left(\, G_{\tau \bar{\tau} , \tau_l} + G_{\tau_l
\bar{\tau} , \tau} -
G_{\tau \tau_l, \bar{\tau}} \, \right) \nonumber \\
& & + \frac{1}{2} \,  G^{\tau_s \tau_m} \left(\, G_{\tau \tau_m ,
\tau_l} + G_{\tau_l \tau_m , \tau} - G_{\tau \tau_l, \tau_m} \,
\right)
\nonumber \\
& & + \frac{1}{2} \,  G^{\tau_s \tau_s} \left(\, G_{\tau \tau_s ,
\tau_l} + G_{\tau_l \tau_s , \tau} - G_{\tau \tau_l, \tau_s} \,
\right),
\nonumber \\
&\sim& \frac{1}{\cal V} \, \frac{1}{{\cal V}^{5/3}} + {\cal V}^{2/3}
\, \frac{1}{{\cal V}^{7/3}} + {\cal V} \,
\frac{1}{{\cal V}^{8/3}}, \nonumber \\
&\sim& \frac{1}{{\cal V}^{5/3}}, \\
{\Gamma^{\tau_s}}_{\tau \tau_s} &=& \frac{1}{2} \, G^{\tau_s
\bar{\tau}} \left(\, G_{\tau \bar{\tau} , \tau_s} + G_{\tau_s
\bar{\tau} , \tau} -
G_{\tau \tau_s, \bar{\tau}} \, \right) \nonumber \\
& & + \frac{1}{2} \,  G^{\tau_s \tau_l} \left(\, G_{\tau \tau_l ,
\tau_s} + G_{\tau_s \tau_l , \tau} - G_{\tau \tau_s, \tau_l} \,
\right) + \frac{1}{2} \,  G^{\tau_s \tau_s} \, G_{\tau_s \tau_s ,
\tau},
\nonumber \\
&\sim& \frac{1}{\cal V} \, \frac{1}{{\cal V}^2} + {\cal V}^{2/3} \,
\frac{1}{{\cal V}^{8/3}} + {\cal V} \,
\frac{1}{{\cal V}^2}, \nonumber \\
&\sim& \frac{1}{\cal V}, \\
{\Gamma^{\tau_s}}_{\tau \bar{\tau}} &=& \frac{1}{2} \, G^{\tau_s
\bar{\tau}} \, G_{\tau \bar{\tau} , \bar{\tau}} + \frac{1}{2} \,
G^{\tau_s \tau_l} \left(\, G_{\tau \tau_l , \bar{\tau}} +
G_{\bar{\tau} \tau_l , \tau} - G_{\tau \bar{\tau}, \tau_l} \,
\right)
\nonumber \\
& & + \frac{1}{2} \,  G^{\tau_s \tau_s} \left(\, G_{\tau \tau_s ,
\bar{\tau}} + G_{\bar{\tau} \tau_s , \tau} - G_{\tau \bar{\tau},
\tau_s} \, \right),
\nonumber \\
&\sim& \frac{1}{\cal V} \, {\cal O}(1) + {\cal V}^{2/3} \,
\frac{1}{{\cal V}^{5/3}} + {\cal V} \,
\frac{1}{{\cal V}^2}, \nonumber \\
&\sim& \frac{1}{\cal V}, \\
{\Gamma^{\tau_l}}_{\tau_m \tau_n} &=& \frac{1}{2} \, G^{\tau_l
\bar{\tau}} \left(\, G_{\tau_m \bar{\tau} , \tau_n} + G_{\tau_n
\bar{\tau} , \tau_m} -
G_{\tau_m \tau_n, \bar{\tau}} \, \right) \nonumber \\
& & + \frac{1}{2} \,  G^{\tau_l \tau_o} \left(\, G_{\tau_m \tau_o ,
\tau_n} + G_{\tau_n \tau_o , \tau_m} - G_{\tau_m \tau_n, \tau_o} \,
\right)
\nonumber \\
& & + \frac{1}{2} \,  G^{\tau_l \tau_s} \left(\, G_{\tau_m \tau_s ,
\tau_n} + G_{\tau_n \tau_s , \tau_m} - G_{\tau_m \tau_n, \tau_s} \,
\right),
\nonumber \\
&\sim& \frac{1}{{\cal V}^{2/3}} \, \frac{1}{{\cal V}^{7/3}} + {\cal
V}^{4/3} \, \frac{1}{{\cal V}^2} + {\cal V}^{2/3} \,
\frac{1}{{\cal V}^{7/3}}, \nonumber \\
&\sim& \frac{1}{{\cal V}^{2/3}}, \\
{\Gamma^{\tau_l}}_{\tau_m \tau_s} &=& \frac{1}{2} \, G^{\tau_l
\bar{\tau}} \left(\, G_{\tau_m \bar{\tau} , \tau_s} + G_{\tau_s
\bar{\tau} , \tau_m} -
G_{\tau_m \tau_s, \bar{\tau}} \, \right) \nonumber \\
& & + \frac{1}{2} \,  G^{\tau_l \tau_n} \left(\, G_{\tau_m \tau_n ,
\tau_s} + G_{\tau_s \tau_n , \tau_m} - G_{\tau_m \tau_s, \tau_n} \,
\right) + \frac{1}{2} \,  G^{\tau_l \tau_s} G_{\tau_s \tau_s ,
\tau_m},
\nonumber \\
&\sim& \frac{1}{{\cal V}^{2/3}} \, \frac{1}{{\cal V}^{8/3}} + {\cal
V}^{4/3} \, \frac{1}{{\cal V}^{7/3}} + {\cal V}^{2/3} \,
\frac{1}{{\cal V}^{5/3}}, \nonumber \\
&\sim& \frac{1}{\cal V}, \\
{\Gamma^{\tau_l}}_{\tau_s \tau_s} &=& \frac{1}{2} \, G^{\tau_l
\bar{\tau}} \left(\, 2 G_{\tau_s \bar{\tau} , \tau_s} -
G_{\tau_s \tau_s, \bar{\tau}} \, \right) \nonumber \\
& & + \frac{1}{2} \,  G^{\tau_l \tau_m} \left(\, 2 G_{\tau_s \tau_m
, \tau_s} - G_{\tau_s \tau_s, \tau_m} \, \right) + \frac{1}{2} \,
G^{\tau_l \tau_s} \, G_{\tau_s \tau_s, \tau_s},
\nonumber \\
&\sim& \frac{1}{{\cal V}^{2/3}} \, \frac{1}{{\cal V}^2} + {\cal
V}^{4/3} \, \frac{1}{{\cal V}^{5/3}} + {\cal V}^{2/3} \,
\frac{1}{\cal V}, \nonumber \\
&\sim& \frac{1}{{\cal V}^{1/3}}, \\
{\Gamma^{\tau_l}}_{\tau \tau_m} &=& \frac{1}{2} \, G^{\tau_l
\bar{\tau}} \left(\, G_{\tau \bar{\tau} , \tau_m} + G_{\tau_m
\bar{\tau} , \tau} -
G_{\tau \tau_m, \bar{\tau}} \, \right) \nonumber \\
& & + \frac{1}{2} \,  G^{\tau_l \tau_n} \left(\, G_{\tau \tau_n ,
\tau_m} + G_{\tau_m \tau_n , \tau} - G_{\tau \tau_m, \tau_n} \,
\right)
\nonumber \\
& & + \frac{1}{2} \,  G^{\tau_l \tau_s} \left(\, G_{\tau \tau_s ,
\tau_m} + G_{\tau_m \tau_s , \tau} - G_{\tau \tau_m, \tau_s} \,
\right),
\nonumber \\
&\sim& \frac{1}{{\cal V}^{2/3}} \, \frac{1}{{\cal V}^{5/3}} + {\cal
V}^{4/3} \, \frac{1}{{\cal V}^{7/3}} + {\cal V}^{2/3} \,
\frac{1}{{\cal V}^{8/3}}, \nonumber \\
&\sim& \frac{1}{\cal V}, \\
{\Gamma^{\tau_l}}_{\tau \tau_s} &=& \frac{1}{2} \, G^{\tau_l
\bar{\tau}} \left(\, G_{\tau \bar{\tau} , \tau_s} + G_{\tau_s
\bar{\tau} , \tau} -
G_{\tau \tau_s, \bar{\tau}} \, \right) \nonumber \\
& & + \frac{1}{2} \,  G^{\tau_l \tau_m} \left(\, G_{\tau \tau_m ,
\tau_s} + G_{\tau_s \tau_m , \tau} - G_{\tau \tau_s, \tau_m} \,
\right) + \frac{1}{2} \,  G^{\tau_l \tau_s} \, G_{\tau_s \tau_s ,
\tau}
\nonumber \\
&\sim& \frac{1}{{\cal V}^{2/3}} \, \frac{1}{{\cal V}^2} + {\cal
V}^{4/3} \, \frac{1}{{\cal V}^{8/3}} + {\cal V}^{2/3} \,
\frac{1}{{\cal V}^2}, \nonumber \\
&\sim& \frac{1}{{\cal V}^{4/3}}, \\
{\Gamma^{\tau_l}}_{\tau \bar{\tau}} &=& \frac{1}{2} \, G^{\tau_l
\bar{\tau}} \, G_{\tau \bar{\tau} , \bar{\tau}} + \frac{1}{2} \,
G^{\tau_l \tau_m} \left(\, G_{\tau \tau_m , \bar{\tau}} +
G_{\bar{\tau} \tau_m , \tau} - G_{\tau \bar{\tau}, \tau_m} \,
\right)
\nonumber \\
& & + \frac{1}{2} \,  G^{\tau_l \tau_s} \left(\, G_{\tau \tau_s ,
\bar{\tau}} + G_{\bar{\tau} \tau_s , \tau} - G_{\tau \bar{\tau},
\tau_s} \, \right),
\nonumber \\
&\sim& \frac{1}{{\cal V}^{2/3}} \, {\cal O}(1) + {\cal V}^{4/3} \,
\frac{1}{{\cal V}^{5/3}} + {\cal V}^{2/3} \,
\frac{1}{{\cal V}^2}, \nonumber \\
&\sim& \frac{1}{{\cal V}^{1/3}}.
\end{eqnarray}


\begin{thebibliography}{999}
\bibitem{gkp} S. B. Giddings, S. Kachru, J. Polchinski, {\it Hierarchies from Fluxes in String Compactifications, Phys. Rev.} {\bf D66} (2002) 106006 [hep-th/0105097].
\bibitem{grana} M. Grana, {\it Flux Compactifications in String
Theory: A Comprehensive Review, Phys. Rept.} {\bf 423} (2006)
91-158 [hep-th/0509003].
\bibitem{racetrack} J. J. Blanco-Pillado {\it et al.}, {\it Inflating in a Better Racetrack}, [hep-th/0603129].
\bibitem{axinfle} R. Holman and J. A. Hutasoit , {\it Axionic
Inflation from Large Volume Flux Compactifications},
[hep-th/0603246].
\bibitem{others} Z. Lalak, G. Ross and S. Sarkar, {\it Racetrack
Inflation and assisted Moduli Stabilisation}, [hep-th/0503178]; A.
Westphal, {\it Eternal Inflation with $\alpha'$ Corrections, JCAP} {\bf 0511} (2005) 003 [hep-th/0507079].
\bibitem{kahler}  J. P. Conlon and F. Quevedo, {\it K\"{a}hler Moduli Inflation, J. High Energy Phys.} {\bf 0601} (2006) 146 [hep-th/0509012]. 
\bibitem{dilatoninflation} R. Brustein and P. Steinhardt, {\it Challenges For Superstring
Cosmology, Phys. Lett. B} {\bf 302} (1993) 196 [hep-th/9212049].
\bibitem{vijay} V. Balasubramanian V, P. Berglund, J. Conlon and F. Quevedo, {\it Systematics of Moduli Stabilization in Calabi-Yau Flux Compactifications, J. High Energy Phys} {\bf 0503} (2005) 007 [hep-th/0502058].
\bibitem{transplanckianme} C. P. Burgess, J. Cline, F. Lemieux, R. Holman,
{\it Are Inflationary Predictions Sensitive to Very High Energy Physics?, J. High Energy Phys.} {\bf 0302} (2003) 048  [hep-th/0210233].
\bibitem{wmap} Spergel {\it et al.},
Wilkinson Microwave Anisotropy Probe (WMAP) Three Year Results:
Implications for Cosmology, http://lambda.gsfc.nasa.gov/product/map
(2006).
\bibitem{becker} K. Becker, M. Becker, M. Haack and J. Louis, {\it Supersymmetry Breaking and $\alpha'$-corrections to Flux
Induced Potentials, J. High Energy Phys} {\bf 0206} (2002) 060
[hep-th/0204254].
\bibitem{GVW} S. Gukov, C. Vafa and E. Witten, {\it CFT's from Calabi-Yau Fourfolds,  Nucl. Phys.} {\bf B584} (2000) 002.
[Erratum-ibid. {\bf B608} (2001) 477] [hep-th/9906070].
\bibitem{inst} E. Witten,  {\it Non-Perturbative Superpotentials in String Theory, Nucl. Phys.}
 {\bf B474} (19916) 343 [hep-th/9604030].
\bibitem{gaugino} C. Burgess, J. Derendinger, F. Quevedo and M. Quir\'{o}s,
{\it Gaugino Condensates and Chiral-Linear Duality: An
Effective-Lagrangian Analysis, Phys. Letter B} {\bf 348} (1995) 
428-442.
\bibitem{denef} F. Denef, M. Douglas and B. Florea, {\it Building a better
Racetrack,  J. High Energy Phys.} {\bf 0406} (2004) 034
[hep-th/0404257].
\bibitem{kklt} S. Kachru, R. Kallosh, A. Linde and S. Trivedi, {\it De Sitter Vacua in String Theory, Phys. Rev. D} {\bf 68} (2003) 046005 [hep-th/0301240].
\bibitem{d-term} C. Burgess, R. Kallosh and F. Quevedo,{\it  de Sitter String Vacua from Supersymmetric
D-terms,  J. High Energy Phys.} {\bf 0310} (2003) 056
[hep-th/0309187].
\bibitem{modkklt} G. Curio and V. Spillner, {\it On the Modified
KKLT Procedure: A Case Study for the $P_{11169}[18]$ Model},
hep-th/0606047.
\bibitem{topology} J. Sim$\acute{\rm o}$n, R. Jimenez, L. Verde, P. Berglund and V. Balasubramanian,
{\it Using Cosmology to constrain to Topology of Hidden Dimensions},
hep-th/0605371.
\bibitem{realistic} C. Burgess, J. Cline, H. Stoica and F. Quevedo,
 {\it Inflation in Realistic D-Brane Models, J. High
Energy Phys.} {\bf 0409} (2004) 033 [hep-th/0403119].
\bibitem{bobkov} K. Bobkov, {\it Volume Stabilization via Alpha'
Corrections in Type IIB Theory with Fluxes, J. High Energy
Phys.} {\bf 0505} (2005) 010 [hep-th/0412239].
\end{thebibliography}
\end{document}